\documentclass{osa-article}

\journal{osajournal}
\articletype{Research Article}
\begin{document}

\title{Double-optical phase-transition in a three level Rydberg state in thermal Rubidium vapor}

\author{Lin Cheng\authormark{1,2}, Kun Huang \authormark{1}, Chunhui Shao\authormark{1}, Fan Wu\authormark{3}, Zhiyuan Xiong\authormark{1}, Yanpeng Zhang\authormark{2}*}

\address{\authormark{1}State Key Laboratory of Widegap Semiconductor Optoelectronic Materials and Technologies, North University of China, Taiyuan, China, 030051\\
}
\address{\authormark{2}Key Laboratory for Physical Electronics and Devices of the Ministry of Education
and Shaanxi Key Lab of Information Photonic Technique, Xi’an Jiaotong
University, Xi’an, Shaanxi 710049, China\\

}

\email{\authormark{*}ypzhang@xjtu.edu.cn} %% email address is required



\begin{abstract}
We report on the observation of electromagnetically induced transparency (EIT) with intrinsic phase transitions in a three-level ladder system within rubidium atomic vapor. The observed abrupt transitions between low and high Rydberg occupancy states manifest in the probe beam transmission, depending on the principal quantum number, the Rabi frequency of the coupling field, atomic density, and probe beam detuning. Our study elucidates the underlying interaction mechanisms governing the EIT phase transition and enriches the existing experiments of multi-parameter regulation phase transitions. These findings establish a robust platform for investigating nonequilibrium phase transitions in atomic ensembles, bridging the gap between classical mean-field theories and microscopic quantum dynamics.
\end{abstract}
%1991Atomic,1999Measurement,The dipole-dipole of highly excited Rydberg states induced level shifts between neighboring states are much larger than the excitation linewidth.
%%%%%%%%%%%%%%%%%%%%%%%%%%  body  %%%%%%%%%%%%%%%%%%%%%%%%%%
\section{Introduction}
The nonlinear response strength of atomic gases is generally several orders of magnitude higher than that of solid materials, which is an ideal platform for studying nonlinear effects~\cite{1995Observation}. When a laser resonates with an atomic transition, it results in a quantum phase between the probe light and the pump light, resulting in asymmetrical absorption lines, i.e., electromagnetically induced transparency,
EIT. In this way, absorption at the frequency of the resonant transition can be eliminated, resulting in a strong nonlinear polarizability accompanied by strong dispersion variations. Rubidium metal atomic vapor excited to high-lying Rydberg states exhibits intrinsic optical phase transition in EIT due to its nonequilibrium dynamics, which has been the subject of several recent studies. EIT provides the control of  population dynamics and enables high resolution probing of the driven-dissipative dynamics. 
Optical bistability means there are two stable output states for the same input parameters, which is a well-studied phenomenon that has provided a rich contribution to the understanding of nonequilibrium systems.  Optical phase transition is a favorable tool for studying non-equilibrium systems, which has been favored by many researchers. Bistability can be found in many materials, such as Fabry-Perot cavities, nonlinear prisms~\cite{1988Bistability}, QED cavities~\cite{2013Femtojoule}, plasmonic nanostructures~\cite{2008Optical}, and liquid crystals~\cite{2014Bistability}. The bistability in these materials was derived from the feedback of the optical cavity~\cite{1987Optical}or cryogenics ~\cite{1994Cooperative}. Moreover, strong interparticle interactions, e.g., dipole-dipole interactions cause the feedback and then induce intrinsic optical bistability as well~\cite{1977Hysteresis}. In the Rydberg states, the dipole-dipole induced level shifts between neighboring states are larger than the excitation linewidth, known as dipole blockade~\cite{M2001Dipole}. Hence, the optical phase transition in the transmission of the probe beam can be observed directly. This phase transition provides a new way to study the optical bistablity~\cite{PhysRevX.10.021023}.

However, there are currently two interpretations, one for the dipole-dipole interaction on the Rydberg state~\cite{2013Femtojoule,2014Bistability}, and the other for the presence of optically phase transition ions in thermal atoms~\cite{2016Charge}. It was not until 2020 that the Ref.~\cite{2020Phase} unified these two methods of interpretation. This work suggests that the shift originates not directly from ionizing collisions but also from avalanche ionization~\cite{2020Phase}.

In this letter, we demonstrate the experimental and theoretical study of the optical phase transition in a thermal Rydberg ensemble. The effects of temperature and different detuning on phase transitions are proposed, which enrich the research on phase transitions. The driving and dissipation within the system can be controlled by  parameters such as power, detuning, atomic density, and quantum number n of the Rydberg state. The experimental results reveal that (1) the increase in the hysteresis window width followed by a subsequent narrowing as n is increased. (2) We also observe a saturation of the bistability width with driving laser intensity. Furthermore, the results are consistent with predictions from a surprisingly simple theoretical model based on the semiclassical Maxwell-Bloch equations including the effect of level shifts and broadening originating from additional processes in the Rydberg manifold. The experimental systems displaying the dynamical phase transitions provide a new way to understand nonequilibrium phenomena.

\section{Experiment Scheme and Basic Theory }

\begin{figure}[h!]
\centering\includegraphics[width=13cm]{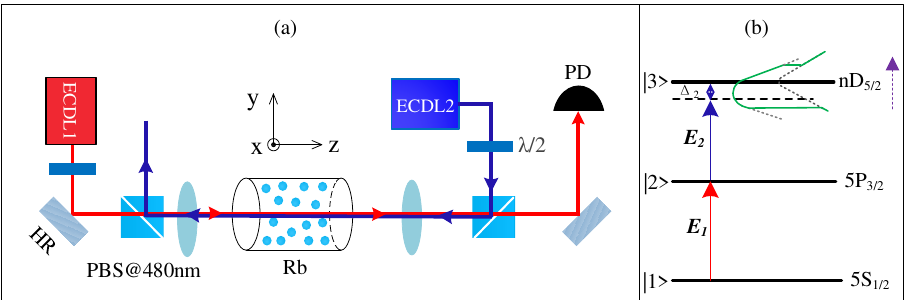}
\caption{(a) Schematic of the experimental setup. The two excitation lasers co-propagate through a 10 cm vapor cell. (b) The ladder configuration with ground state and Rydberg state $nD_{1/2}$ ($n=37,45,54$). The levels are coupled by a laser with Rabi frequency $\Omega_i$ and detuning from resonance $\Delta_i$.}
\label{Fig1}
\end{figure}  

The experimental setup is shown schematically in Fig.\ref{Fig1}(a). A collimated Gaussian beam profile with a diameter of 0.5 mm is obtained by the beam from an external cavity diode laser (ECDL) at a center wavelength of $\sim$ 780 nm as the probe beam. The probe field couples the $^{85}Rb 5S_{1/2}(|1\rangle)\rightarrow{5P_{3/2}}(|2\rangle)$ transition and then is detected at a photodiode, while the coupling field with the frequency-stabilized 480 nm drives the $ 5P_{3/2}(|2\rangle)\rightarrow{nD_{1/2}}(|3\rangle)$. Here, the $\bf{E_1}$ (780 nm) beam is locked on resonance and $\bf{E_2}$ (480 nm) have a variable detuning $\Delta_2$ around the resonance position. $\Gamma_{2(3)}$ are the decay rates of $|2\rangle$ and $|3\rangle$, with the $\Gamma_2<<\Gamma_3$. Two laser beams $\bf{E_i}$, coming from two excited cavity diode lasers (ECDLs), initially are counterpropagating and converge at the center of thermal Rb cell. The Rb cell have a length of 10 cm long, which is maintained at the temperature of 80$^\circ$C giving a number density, $N_0$ of $1.5*10^{13}$cm$^{-3}$. The parameters of laser beams are defined as follows: frequency detuning $\Delta_i=\omega_i-\Omega_i$ denotes the difference between the frequency $\omega_i$ of $\bf{E_i}$  and the resonant transition frequency $\Omega_i$. $G_i=\mu_{jk}E_i/h$ is the Rabi frequency between the energy |j〉 and |k〉 with $\mu_{jk}$ being the transition dipole matrix element. We access transitions to $nP_{3/2}$ states in the range $n=37D_{5/2}, 45D_{5/2}, 54D_{5/2}$, corresponding the center wavelength of the dressing field $\bf{E_2}$ is about 481.0487 nm, 480.3850 nm, 479.9698 nm. The Rydberg laser is tuned around the resonance between the state $5P_{1/2}$ and the Rydberg states $nD_{5/2}$. Through scanning $\Delta_2$, the spectra of $\bf{E_1}$ field can be obtained by the oscilloscope. Distinct transmission with phase transition optical response can be obtained when we change the multi-parameters.

 We model the system using the density-matrix applied in the Ladder system, as shown in Fig.\ref{Fig1}(b). We consider a ground state $5P_{1/2}$ and a Rydberg state $nD_{5/2}$ coupled by a laser with Rabi frequency $\Omega_2$ and detuning $\Delta_2$. A strong dynamical nonlinearity comes from the avalanches interaction between Rydberg states. The phase transition is triggered by a self-organized criticality. Using semiclassical analysis, the time evolution of the system is described by a Lindblad master equation applied to a Ladder system. We use threshold modified mean-field theory~\cite{PhysRevX.10.021023} to describe the dipole-dipole interaction between the Rydberg atom in a classical approximation, where the many-body interaction is described in terms of the response of a single atom interacting with a mean-field interaction potential. In the Rydberg states, there are two thermodynanical phases around  a critical $N_{R,(c)}$: non-interacting phase (NI phase) with Lydberg density $N_R<N_{R,(c)}$ and strong interacting phase (I phase) with $N_R>N_{R,(c)}$~\cite{2020Phase}. The phase transition is induced by the avalanches that relate to either dipole-dipole interaction or ionization collisions with electrons, ions or other atoms. This results in a transition frequency 
 \begin{equation}
\Delta_2\rightarrow\Delta_{eff}=\Delta_2+\Delta'_{(N_R)}
\label{1}
 \end{equation}
 where $\Delta_{eff}$ is the effective detuning~\cite{2012Collective}. $\Delta'_{(N_R)}$ is the mean-field shift (additional detuning) expressed as $\Delta'_{(N_R)}=V\times\rho_{33}=\alpha{N_R}$ when $N_R$ above $N_R,(c)$. Here, $\rho_{33}$ is the population of the Rydberg state and V is the dipole-dipole interaction term over the excitation volume~\cite{2016Intrinsic}.  The enhanced decay $\Gamma$ model should be modified to 
 \begin{equation}
\Gamma_{13}\rightarrow\Gamma_{eff}=\Gamma_{13}+\Gamma'_{N_R}=\Gamma_{13}+\beta{N_R}
\label{2}
 \end{equation}
 where $\Gamma'$ is the enhanced decay which is proportional to the Rydberg density $N_R$ above $N_{R,(c)}$. $\beta$ is the interaction-broadening coefficient and $N$ is the atomic density. The $\beta$ is related to the ionization~\cite{1982Penning} or photoionization~\cite{1994Rydberg}, collisional process~\cite{2008Rydberg} that we cannot distinguish each contribution exactly.
Considering the Doppler effect, $\delta=\Delta_1+\Delta_2\rightarrow\delta+\Delta_{D},\Delta_1\rightarrow\Delta_1+\frac{\omega_1v}{c},\Delta_2\rightarrow\Delta_2-\frac{\omega_2v}{c}$.  $\Delta_D=(\omega_1-\omega_2)v/c$ denotes Doppler shift. 

\begin{table}[t]
\centering\includegraphics[width=13cm]{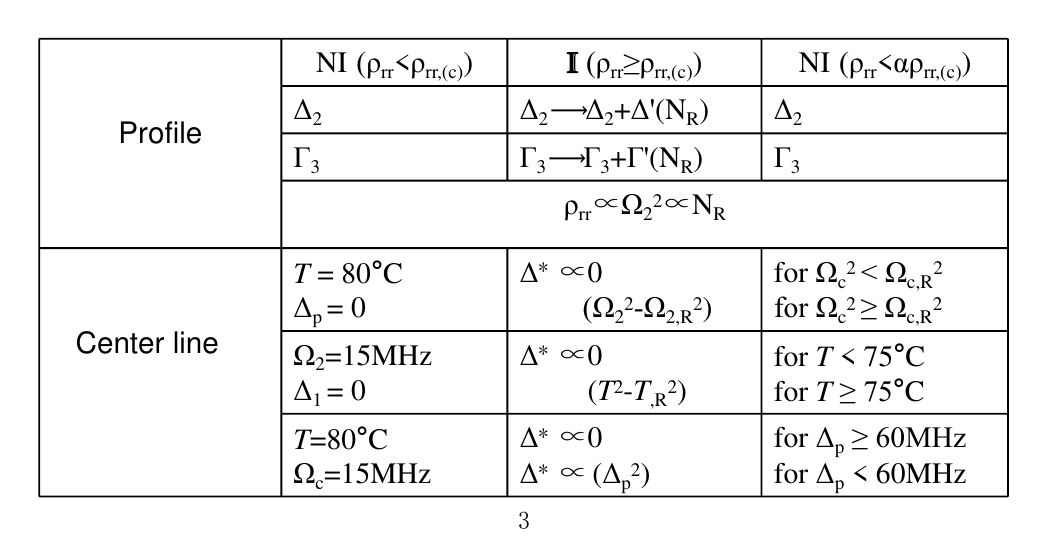}
\caption{The parameters in NI and I phase}
\label{table:1}
\end{table}
Hence, the complex susceptibility of the EIT including the Doppler effect due to the thermal atomic motion is 
 \begin{equation}
\chi_{(v)}=\frac{2N|\mu_{21}|}{\epsilon_0\epsilon_1}\rho_{21}
\label{3}
 \end{equation}
 with 
% \begin{equation}
%\rho_{12}=\frac{(2i(\delta+\Delta_D)-2\gamma_{13})\Omega_p}{(2\Delta_p+2\omega_pv/c+2i\gamma_{12})(2i(\delta+\Delta_D)-2\gamma_{13})-i\Omega_c^2}.
  % \end{equation} 
\begin{equation}
\rho_{21}=\frac{iG_1}{2[\gamma_{12}+i(\Delta_1+\omega_1v/c)]+\frac{|G_2|^2}{2[\gamma_{13}+i(\delta+\Delta_D)]}}
\label{4}
\end{equation}
where $N=N_0e^{-v^2/V_p^2}/\sqrt{\pi}v_p$ represents the distributed Maxwell distribution of the velocity of atoms in a thermal atom ensemble, and the most probable velocity is $v_p=\sqrt{2k_BT/m}$.
The transmission through the EIT medium can be obtained from the susceptibility via
 \begin{equation}
T\sim{e}^{-Im[kL\chi(v)dv]}
\label{5}
 \end{equation}

\section{Results and Discussion}

\begin{figure}[htb!]
\centering\includegraphics[width=10cm]{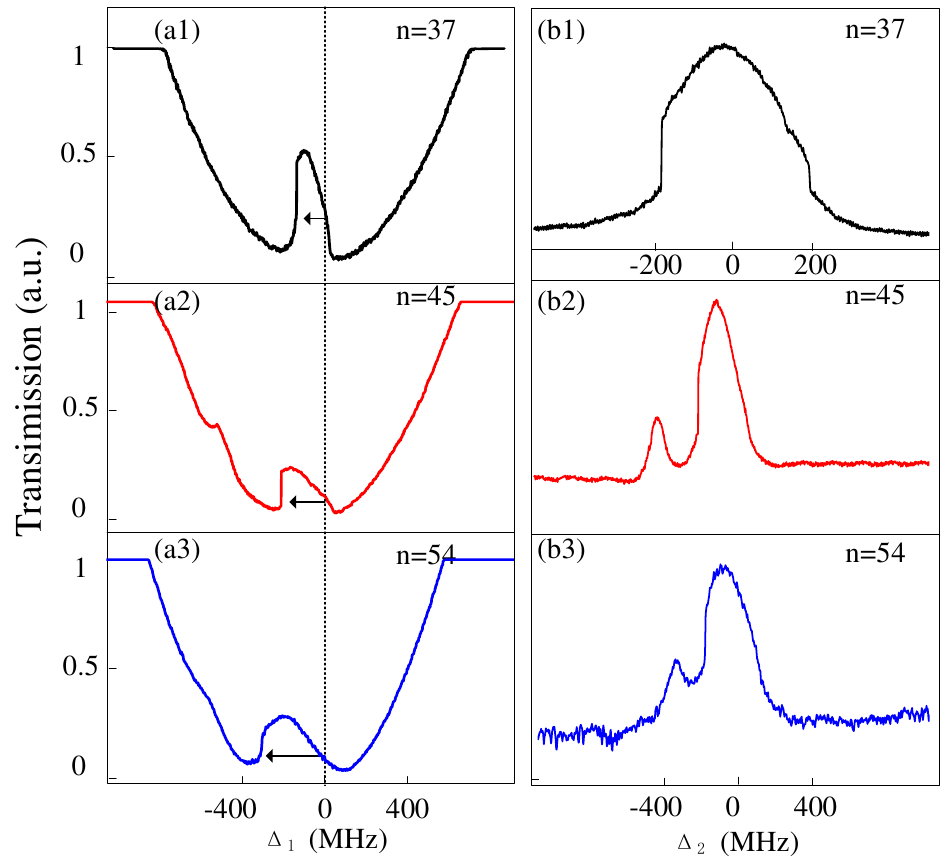}
\caption{(a1)-(a3) The transmission for a resonant probe while scanning the probe beam {$\Delta_1$} at $n=37$, 45 and 54, respectively. (b1)-(b3) The transmission for a resonant probe while scanning the Rydberg laser detuning {$\Delta_2$} at $n=37$, 45 and 54, respectively.}
\label{Fig2}
\end{figure}
Fig. \ref{Fig2} shows the typical phase transition of EIT as a function of the probe detuning for different Rydberg states.  As the level of Rydberg population increases, the excitation-dependent shift produces an asymmetry in the line shape. It was found that as the principal quantum number of the Rydberg state increased, the EIT width also increased before going through a maximum. Eventually, when the shift is greater than the linewidth, the line shape exhibits phase transition. The frequency shift of the phase transition is seen to scale with the fourth power of the principal quantum number ($\sim n^4$) due to the interaction potential between dipole-dipole is $V_{dd}\propto d^2$. Here $d$ is the dipole moment, which is proportional to $n^2$. 

In the three-level EIT medium, Rydberg-mediated phase transitions are observed. %Scanning probe beam alters the Rydberg population and thus mean interaction strength as $N_R\rho_{rr}G_p^2$ as shown in Fig.\ref{Fig2}(a).
However,  scanning coupling beam can obtain EIT transmission spectra with the probe beam on resonance, as shown in Fig. \ref{Fig2}(b). The phase transitions of the EIT still exist, but because they are too close to the other neighboring Rydberg states, the EIT is connected to each other. So we study the effect of Rabi frequency on the phase transition of D=37 separately through scanning the coupling laser subsequently. 

\begin{figure}[htb!]
\centering\includegraphics[width=10cm]{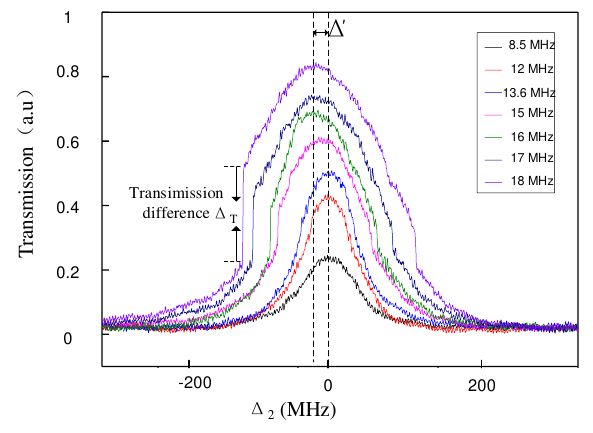}
\caption{Deformation of the EIT transmission peak towards phase transition for different Rabi frequency of coupling beam as $\Delta_2$ is scanned from blue to red detuing when $P_1=0.67$ mW. $\Delta{T}$ is the phase transition, and $\Delta^*$ is the spectral shift.
}
\label{Fig3}
\end{figure}
 When the coupling laser is scanned instead, a similar lineshape is observed but on a flat transmission background as shown in Fig. \ref{Fig3}. The EIT is induced in a narrow frequency window around two-photon resonance $\Delta_1+\Delta_2=0$. Here, the transmission is on resonance $\Delta_1=0$. The black line indicates that the system is in the NI phase at $G_2=8.5$ MHZ and 12 and 13.6 MHz. When the frequency is scanned from the red detuned to the blue,  the traces of the EIT as the increasing Rabi frequency of the coupling laser, the EIT peak is shifted to the red and becomes more and more distorted with the formation of EIT from symmetry to asymmetry accompanying as shown in Fig.\ref{Fig3}.  As the increases of $G_2$ to 15 MHz, a sudden drop in the transmission can be observed  indicating the transition from the NI to the I phase. With the increase of Rabi frequency, on the one hand, the amplitude of EIT increases due to higher $\rho_{21}$; On the other hand, the maxima of EIT shift $\Delta^*$ towards red due to the $N_R\propto\rho_{33}\propto{G}_2^2$. This phenomenon is the result of a competition between a nonlinear energy shift due to an interaction effect, which is dependent on the Rydberg state. There is a buildup of Rydberg population that sustains the ability to excite Rydberg atoms even away from resonance. When the detuning becomes too large, the decay mechanisms prevail and the population breaks down.  It is worth noting that the phase transitions appear on both sides of the resonance. The I phase part presents a broadened EIT due to the enhanced decay $\Gamma'$. In I phase, the sufficient population has been created and it continues to be maintained. The phase transition on the right of EIT is farther from the center frequency than the phase transition on the left which results from the $\Delta'$ and implies an redshift of $|3\rangle$.

%Figure 2 shows the traces of the EIT as the increasing Rabi frequency of the coupling laser, the EIT peak is shifted to the red and becomes more and more distorted, until at sufficiently large intensity the system becomes bistable. [8][as in Ref9], 
%[图2 其中一个随着失谐的变化图]
%[%图3 其中一个随着功率的变化图]
%[图4 其中一个随着温度的变化图]
%[图2不同n的宽度对比图]
%[两种方案的理论罗列出来，分布模拟与结果对照]
%[文章创新的：1 不同n态的验证 2 验证里德堡双稳的理论符合哪种 3 找其它不同点]

\begin{figure}[ht!]
\centering\includegraphics[width=11cm]{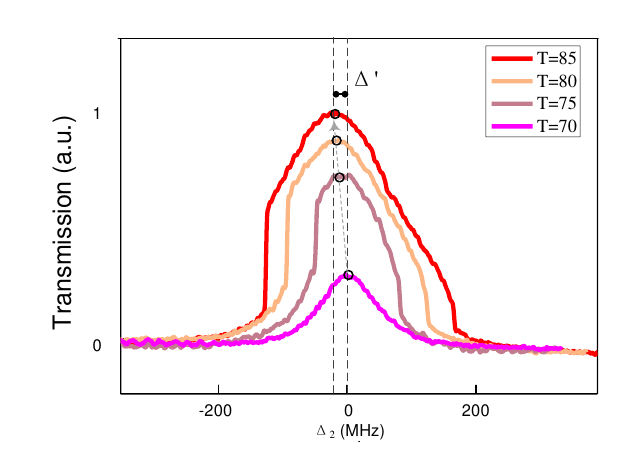}          
\caption{the deformation of the EIT for the increasing the temperature. (a) from the 70-85$^{\circ}$C  at intervals of 5$^{\circ}$C.}
\label{Fig4}
\end{figure}

In Fig.\ref{Fig4},  the atomic density $N_0$ increases as the temperature goes up, expressed as $lg N_0=4.312-4040/T-lg(kT)$~\cite{Lin2019Manipulation}. On the one hand, the amplitude of EIT increases; On the other hand, the maxima of EIT all shift towards shorter wavelengths, as shown in Fig.\ref{Fig4}.  The transmission of the probe beam resonant with the optical transition is increased by the population in the Rydberg state and atomic density $N_0$ in Eq.\eqref{3}. The increase of temperature directly causes the increase of population, and in turn, phase transitions are observed without increasing the power of the coupling field. As the Rydberg population increases, the excitation-dependent shift first produces an asymmetry EIT when  $T$ approaches 75${^\circ}$.  Eventually, when the shift is greater than the width of the optical resonance, the EIT exhibits optical phase transition  at $T=75^\circ$.

\begin{figure}[ht!]
\centering\includegraphics[width=10cm]{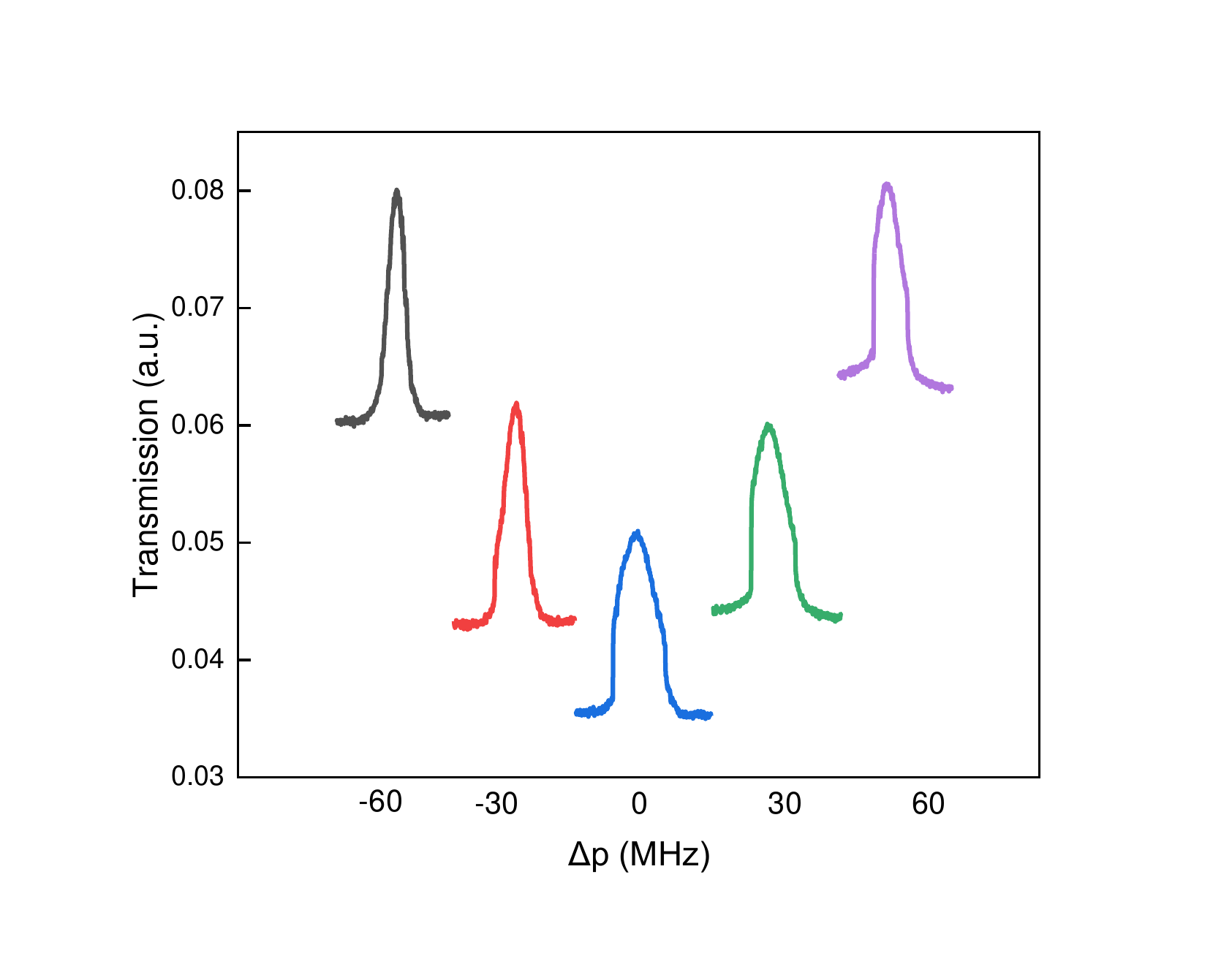}
\caption{ Observation of optical bistability in EIT transmission spectra. Transmission of probe light as $\Delta_2$ is scanned from red to blue detuning. As the voltage increases from left to right, the wavelength becomes smaller and the frequency increases.}
\label{Fig5}
\end{figure}
In previous study, we studied the multi-parameters controlling the threshold of the phase transition without changing the probe beam detuning. Unlike the Ref~\cite{2020Phase}, if the detuning is not appropriate at certain $G_1$, the phase transition cannot be observed. Optical transition is observed on both sides of the EIT when we scan the coupler beam in different $\Delta_1$. The entire profile presents an absorption dip expressed as $\rho_{21}=iG_2/(i\Delta_2+\Gamma_{21})$. When different probe detuned, the resulting EIT position depends on the detuning position of the probe. When we turn on the $E_2$, the absorption dip of $E_1$ appears with annn EIT peak. The positions of the EIT peak can be moved by changing the $\Delta_1$ at $\Delta_1+\Delta_2=0$. When $\Delta_1=-60$ MHz, the coupling field is scanned, and the normal EIT phenomenon occurs. When $\Delta_1=-30$ MHz, the coupling field is scanned, and the right side of the EIT begins to show a phase transition. When $\Delta_1=0$, the left side of the EIT also has a phase transition, and the $\Delta_1$ value continues to be adjusted to $\Delta_1=30$ and 60 MHz, and the phase transition does not disappear.
The change in probe laser transmission as a function of Rydberg detuning is shown in Fig. \ref{Fig5}. As the level of Rydberg population increases, on both sides of the resonance, the phase transition becomes unilateral with initially an ordinary EIT.
Hence, the Rydberg density $N_R$ depends not only on the coupling field but also on the detuning of the probe beam.

\section{Conclusion}
In summary, we investigated  optical phase transition in the Rydberg state of Rb vapor in a ladder system. We characterized the behavior of the FWHM as a function of the principal quantum number $n$. Moreover, the phase transition can be modulated not only by the power of the coupling beam, atomic density, but also by the frequency detuning of the probe beam. Multi-parameters influence the energy shift and decay rates. This work provides a complementary experiment that showed that phase transitions can also be affected by the frequency detuning of the probe beam for the intrinsic optical phase transition and a rich contribution for understanding the nonequilibrium systems. 
\section{Funding}
{We would like to acknowledges the support the National Natural Science Foundation of China (62305312), the Natural Science Foundation of Shanxi Province, China (202203021222021), Research Project Supported by Shanxi Scholarship Council of China(2312700048MZ) and the fellowship of China Postdoctoral Science Foundation (2022M722923). Shanxi Provincial Teaching Reform and Innovation Project (2024YB008). }

\bibliography{sample}

\begin{thebibliography}{10}
\newcommand{\enquote}[1]{``#1''}

\bibitem{1995Observation}
Y.~Q. Li and M.~Xiao, \enquote{Observation of quantum interference between dressed states in an electromagnetically induced transparency,} {\protect\JournalTitle{Physical Review A}} \textbf{51}, 4959 (1995).

\bibitem{1988Bistability}
G.~I. Stegeman, G.~Assanto, R.~Zanoni, C.~T. Seaton, E.~Garmire, A.~A. Maradudin, R.~Reinisch, and G.~Vitrant, \enquote{Bistability and switching in nonlinear prism coupling,} {\protect\JournalTitle{Applied Physics Letters}} \textbf{52}, 869--871 (1988).

\bibitem{2013Femtojoule}
Y.~D. Kwon, M.~A. Armen, and H.~Mabuchi, \enquote{Femtojoule-scale all-optical latching and modulation via cavity nonlinear optics,} {\protect\JournalTitle{Physical Review Letters}} \textbf{111}, 203002 (2013).

\bibitem{2008Optical}
F.~Y. Wang, G.~X. Li, H.~L. Tam, K.~W. Cheah, and S.~N. Zhu, \enquote{Optical bistability and multistability in one-dimensional periodic metal-dielectric photonic crystal,} {\protect\JournalTitle{Applied Physics Letters}} \textbf{92}, 213903 (2008).

\bibitem{2014Bistability}
N.~Kravets, A.~Piccardi, A.~Alberucci, O.~Buchnev, and G.~Assanto, \enquote{Bistability with optical beams propagating in a reorientational medium,} {\protect\JournalTitle{Phys.rev.lett}} \textbf{113}, 023901 (2014).

\bibitem{1987Optical}
H.~M. Gibbs and D.~Sarid, \enquote{Optical bistability: controlling light with light,} {\protect\JournalTitle{Physics Today}} \textbf{40}, 71--72 (1987).

\bibitem{1994Cooperative}
M.~P. Hehlen, H.~U. Güdel, Q.~Shu, J.~Rai, S.~Rai, and S.~C. Rand, \enquote{Cooperative bistability in dense, excited atomic systems,} {\protect\JournalTitle{Phys.rev.lett}} \textbf{73}, 1103 (1994).

\bibitem{1977Hysteresis}
H.~J. Carmichael and D.~F. Walls, \enquote{Hysteresis in the spectrum for cooperative resonance fluorescence,} {\protect\JournalTitle{Journal of Physics B Atomic and Molecular Physics}} \textbf{10}, L685 (1977).

\bibitem{M2001Dipole}
M., D., Lukin, M., Fleischhauer, R., Cote, L., M., and Duan, \enquote{Dipole blockade and quantum information processing in mesoscopic atomic ensembles,} {\protect\JournalTitle{Physical Review Letters}} \textbf{87}, 37901--37901 (2001).

\bibitem{PhysRevX.10.021023}
D.-S. Ding, H.~Busche, B.-S. Shi, G.-C. Guo, and C.~S. Adams, \enquote{Phase diagram and self-organizing dynamics in a thermal ensemble of strongly interacting rydberg atoms,} {\protect\JournalTitle{Phys. Rev. X}} \textbf{10}, 021023 (2020).

\bibitem{2016Charge}
D.~Weller, A.~Urvoy, A.~Rico, R.~Lw, and H.~Kübler, \enquote{Charge-induced optical bistability in thermal rydberg vapor,} {\protect\JournalTitle{Physical Review A}}  (2016).

\bibitem{2020Phase}
D.~S. Ding, H.~Busche, B.~S. Shi, G.~C. Guo, and C.~S. Adams, \enquote{Phase diagram and self-organizing dynamics in a thermal ensemble of strongly interacting rydberg atoms,} {\protect\JournalTitle{Physical Review X}}  (2020).

\bibitem{2012Collective}
T.~E. Lee, H.~H?Ffner, and M.~C. Cross, \enquote{Collective quantum jumps of rydberg atoms,} {\protect\JournalTitle{Physical Review Letters}} \textbf{108}, 23602--23602 (2012).

\bibitem{2016Intrinsic}
N.~R. De~Melo, C.~G. Wade, N.~Sibalic, J.~M. Kondo, C.~S. Adams, and K.~J. Weatherill, \enquote{Intrinsic optical bistability in a strongly-driven rydberg ensemble,} {\protect\JournalTitle{Physical Review A}}  (2016).

\bibitem{1982Penning}
M.~Cheret, L.~Barbier, W.~Lindinger, and R.~Deloche, \enquote{Penning and associative ionisation of highly excited rubidium atoms,} {\protect\JournalTitle{Journal of Physics B}} \textbf{15}, 3463---34773477 (1982).

\bibitem{1994Rydberg}
T.~Gallagher, \enquote{Rydberg atoms,} {\protect\JournalTitle{Cambridge University Press}}  (1994).

\bibitem{2008Rydberg}
A.~Reinhard, T.~C. Liebisch, K.~C. Younge, P.~R. Berman, and G.~Raithel, \enquote{Rydberg-rydberg collisions: Resonant enhancement of state mixing and penning ionization,} {\protect\JournalTitle{Physical Review Letters}} \textbf{100}, 123007 (2008).

\bibitem{Lin2019Manipulation}
Lin, Cheng, Zhaoyang, Zhang, Lei, Danmeng, Ma, Gaoguo, Yang, and Tian, \enquote{Manipulation of a ring-shaped beam via spatial self- and cross-phase modulation at lower intensity.} {\protect\JournalTitle{Physical Chemistry Chemical Physics Pccp}}  (2019).

\end{thebibliography}
\end{document}